\documentclass[12pt]{article}
\begin{document}
\newcommand{\ben}{\begin{enumerate}}

\newcommand{\een}{\end{enumerate}}

\newcommand{\be}{\begin{equation}}

\newcommand{\ee}{\end{equation}}

\newcommand{\bse}{\begin{subequation}}

\newcommand{\ese}{\end{subequation}}

\newcommand{\bea}{\begin{eqnarray}}

\newcommand{\eea}{\end{eqnarray}}

\newcommand{\bc}{\begin{center}}

\newcommand{\ec}{\end{center}}

\newcommand{\mb}{\mbox{\ }}

\newcommand{\vs}{\vspace}

\newcommand{\ra}{\rightarrow}

\newcommand{\la}{\leftarrow}

\newcommand{\IR}{\mbox{I \hspace{-0.2cm}R}}

\newcommand{\IN}{\mbox{I \hspace{-0.2cm}N}}

\newcommand{\ol}{\overline}

\newcommand{\ul}{\underline}

\newcommand{\ts}{\textstyle}

\author        {
       {\bf  Ay\c{s}e H\"{u}meyra B\.ILGE}\\
        {\it  Department of Mathematics, Istanbul Technical University}\\
        {\it  Istanbul, Turkey}\\
        {\it  e-mail: bilge@itu.edu.tr}\\
        {\bf  Tekin DEREL\.I}\\
        {\it  Department of Physics, Ko\c{c} University}\\
        {\it  Istanbul, Turkey}\\
        {\it  e-mail: tdereli@ku.edu.tr}\\
        {\bf  \c{S}ahin KO\c{C}AK}\\
        {\it  Department of Mathematics, Anadolu University}\\
        {\it  Eski\c{s}ehir, Turkey}\\
        {\it  E.mail : skocak@anadolu.edu.tr}\\
        }

\title{Seiberg-Witten Type Monopole Equations \\ on 8-Manifolds with $Spin(7)$
Holonomy \\as Minimizers of a Quadratic Action }

\maketitle

\begin{abstract}
\noindent We obtain an elliptic system of monopole equations on
8-manifolds with Spin(7) holonomy
by minimizing an action involving negative spinors coupled to an
Abelian gauge field. 
\end{abstract}

\newpage

\section{Introduction}

In Seiberg-Witten theory over a 4-manifold, a $U(1)$ connection is
coupled to a spinor field.
The Seiberg-Witten equations are the minimizers of
an action involving the curvature 2-form $F$
and a positive Dirac spinor $\phi^+$ (see Eq.(2.7)),
$$
I(A,\phi^+) =\int_M
\left[ \vert F\vert^2
+\vert \nabla \phi^+ \vert^2
+\ts \frac{1}{4}s \vert \phi^+\vert^2
+\ts \frac{1}{4}  \vert \phi^+ \vert^4
\right] {\rm dvol}
\ge \int_M {\rm tr}F^2,
\eqno(1.1)$$
where $s$ is the scalar curvature of the 4-manifold $M$
 and $A$ and $F$ are respectively the connection and
curvature of a line bundle associated
with the spin$^c$ structure on $M$ \cite{Morgan}.
In the coupling of a Dirac spinor to the Yang-Mills action, the key is the
Weitzenb\"ock formula which relates the covariant derivatives $\nabla \phi^\pm$
to $D^\pm \phi^\pm$, where $D^\pm$ are Dirac operators
(see Eqs.(2.15-17)), compensating for the scalar curvature term and
bringing in the  coupling $(\rho^\pm(F)\phi^\pm,\phi^\pm)$.

In 4 dimensions, $\rho^+(F)$ is a $2\times 2$ traceless, hermitian matrix
such that  $\rho^+(F^-)$ is identically zero.  If $D^+\phi^+=0$, the action
(1.1) can be written as
$$I(A,\phi^+)=\int_MF\wedge F +\int_M
\left[ 2\vert F^+\vert^2
-(\rho^+(F^+)\phi^+,\phi^+)
+\ts \frac{1}{4}  \vert \phi^+ \vert^4
\right] {\rm dvol}
\eqno(1.2)$$
Writing
$$
\rho^+(F^+)=\pmatrix{a & b+ic \cr b-ic & -a\cr},\quad
\phi^+=\pmatrix{\phi_1\cr \phi_2\cr},\eqno(1.3)$$
and using $2\vert F^+\vert^2=\vert \rho^+(F^+)\vert^2$, the second integrand in
Eq.(1.2) is written as
\begin{eqnarray}
2\vert F^+\vert^2&-&(\rho^+(F^+)\phi^+,\phi^+)+\ts\frac{1}{4}
   \vert \phi^+\vert^4=
   [a^2+b^2+c^2]\nonumber\\
   & -&[a(\phi_1\bar{\phi}_1-\phi_2\bar{\phi}_2)
                 +b(\phi_1\bar{\phi}_2+\phi_2\bar{\phi}_1)
               +ic(-\phi_1\bar{\phi}_2+\phi_2\bar{\phi}_1)]\nonumber\\
& +&\ts\frac{1}{4}[ \phi_1^2\bar{\phi}_1^2
                    +2\phi_1\bar{\phi}_1\phi_2\bar{\phi}_2
                    + \phi_2^2\bar{\phi}_2^2]  \nonumber 
\end{eqnarray}
$$
=   [a-\ts \frac{1}{2}(\phi_1\bar{\phi}_1-\phi_2\bar{\phi}_2)]^2
     + [b-\ts \frac{1}{2}(\phi_1\bar{\phi}_2+\phi_2\bar{\phi}_1)]^2
     + [c-\ts \frac{i}{2}(-\phi_1\bar{\phi}_2+\phi_2\bar{\phi}_1)]^2.
\eqno(1.4)$$
It can be seen that the vanishing of this term gives the second part of the
Seiberg-Witten equations
$$
D^+\phi^+=0,\quad \rho^+(F^+)=[\phi^+(\bar{\phi}^+)^t]_o,\eqno(1.5)
$$
where the subscript $o$ denotes the trace-free part of a matrix.

From Eq.(1.4), it is clear that the  term
$\frac{1}{4}\vert \phi^+\vert^4$ is added just to complete the square.
Choosing $\rho^+(F^+)$
as in Eq.(1.5), reduces the
action to its topological lower bound.
In a sense, the spinors $\phi^\pm$ are chosen
in such a way
that the coupling is independent of the  part of $F$ of which it is expected to be free.
This will be our idea in looking for Seiberg-Witten type monopole equations
on 8-manifolds.

If one expects to get linear field equations as absolute minimizers of an
action, the
topological  lower bound
for the Yang-Mills functional would be an integral of the second Chern class
$c_2$,
which is proportional to $tr(F^2)$.  For an arbitrary 8-manifold, there is no
canonical way of obtaining a topological invariant, either by  integrating
$c_2$ over a submanifold,
or by multiplying $c_2$ with an appropriate (dimensionless) 4-form and
integrating over $M$. Nevertheless, for  eight  manifolds with
Spin(7) holonomy \cite{Joyce}, there
is a globally defined 4-form, called the {\sl Bonan
4-form} \cite{Bonan}, \cite{Harvey}, which  determines a topological
 lower bound
for the Yang-Mills action 
$$\int_M \vert F\vert^2{\rm  dvol }\ge \int_M F\wedge F\wedge\Phi.\eqno(1.6)$$
The Bonan 4-form  $\Phi$
 belongs to
the class of forms called {\sl calibration forms} \cite{Lawson}.
The 4-form $\Phi$ acting on 2-forms defines a ``$\Phi$-duality" by Eq.(A.2)
(see Appendix A).
This determines
a splitting of the curvature  2-form $F$ into 7 and 21 dimensional
subspaces as
$F=F^{(7)}+F^{(21)}.$
The requirement that $F^{(7)}=0$ leads to a set of 7
equations, proposed originally by Corrigan et.al, \cite{Corrigan}. 
 Together with the Coulomb gauge condition, these equations from an elliptic system
\cite{Bilge}. Gauge theories in higher dimensions in general were  studied 
in \cite{Donaldson}, \cite{Baulieu}, \cite{Acharya}.
The coupling of a spinor field to the gauge fields in
8-dimensions
may follow different paths. A direct generalization of the Seiberg-Witten
theory in 4-dimensions to 8-dimensions  has led to trivial
results \cite{BDK1}. In our previous paper \cite{BDK2}, we defined a 
generalization of the monopole equations by interpreting the right hand sides 
of the Seiberg-Witten equations as a
projection onto a subspace determined by the map $\rho^+$ and coupled
positive spinors to the curvature.  However, it was not possible to
express these equations as absolute minimizers of an action.
Another set of monopole equations in 8-dimensions is proposed by Gao and Tian \cite{Tian}
with the suggestion that it may provide  the S-dual of non-Abelian instantons \cite{Witten}.

In the present paper, we define monopole equations  via a
projection by the map $\rho^-$, thus coupling the curvature to negative
spinors.  This new set of monopole equations are absolute minimizers of an
action, provided that the real and imaginary parts of the negative spinor
belong to  certain complementary subspaces of
$W^-$ determined by $\rho^-(\Phi)$.

\section {Preliminaries.}
\vskip 0.2 cm
\noindent
{\bf Notation.}
We define a Hermitian inner product on $Hom(C^n,C^m)$ by
$$(A,B)=\textstyle {\frac{1}{n}}  tr (\bar{A}^tB).\eqno(2.1)$$
for $A$, $ B$ in $Hom(C^n,C^m)$.
In particular, if $X,Y\in C^n\simeq Hom(C,C^n)$, and $A\in End(C^n)$,
$$(AX,Y)=tr(\bar{X}^t\bar{A}^tY)=
tr(\bar{A}^tY\bar{X}^t)=
n(A,Y\bar{X}^t).\eqno(2.2)$$
Furthermore if $A$ is skew-symmetric, then
$$(AX,X)=\textstyle\frac{n}{2} (A,X\bar{X}^t-\bar{X}X^t).\eqno(2.3)$$
The pointwise norm of a real differential form
$\omega\in \Lambda^*(T_p^*M)$ is defined as
$$\vert
\omega\vert^2=(\omega,\omega) = *({\omega}\wedge*\omega)\eqno(2.4)$$
where $*$ denotes Hodge dual.
We also note that if $F=i\omega$, where $\omega $ is real, then
$$(F,F)=(\omega,\omega),\quad F\wedge F=-\omega\wedge *\omega.\eqno(2.5)$$

\medskip
\noindent
{\bf Spin$^c$ structures.}
A detailed summary of the spin$^c$ structures was given in \cite{Morgan}. We remind that
a spin$^c$-structure on a $2n$-dimensional real inner-product
space $V$ is
a pair $(W,\Gamma)$, where $\Gamma$ is a $2^n$ dimensional hermitian vector
space and $\Gamma:V\to End(W)$ is a linear map
satisfying
$$
{\Gamma(v)}^*=-\Gamma(v),\qquad {\Gamma(v)}^2=-{\vert v \vert}^2
\eqno(2.6)
$$
for $v\in V$.
Globalization of this construction on each fiber to the tangent bundle of a
 $2n$-dimensional
oriented manifold $M$ will be possible if  and only if
 the second Stiefel-Whitney class
$w_2(M)$ has an integral lift.
The spin$^c$ structure on the manifold $M$ will  then be given by the map
 $\Gamma :
 TM \rightarrow End(W)$, where $W$ is now a $2^n$-dimensional
complex Hermitian vector bundle on $M$.
The $\pm i^n$ eigenspaces of $\Gamma(e_{2n}e_{2n-1}\dots e_1)$ determines a
splitting
$$W=W^+\oplus W^-,\eqno(2.7)$$
and the elements of $W^+$ and $W^-$ are called respectively ``positive" and
``negative" spinors.

Restriction of $\Gamma(v)$ to $W^-$ determines a linear map
$\gamma :V\rightarrow Hom(W^-,W^+)$ which satisfies
$\gamma(v)^*\gamma(v)={\vert v\vert}^2$ for all $v\in V$.
The linear map can be recovered from $\gamma$ via
$$\Gamma(v)=\pmatrix{0&\gamma(v)\cr-\gamma(v)^*&0\cr}.\eqno(2.8)$$
To define a spin$^c$ structure on $V$ it is thus enough to give a linear map
$\gamma :V\rightarrow Hom(W^-,W^+)$ satisfying
$\gamma(v)^*\gamma(v)={\vert v\vert}^2$ for all $v\in V$.

In our case of an 8-manifold $M$ with Spin(7)-holonomy, we will instantly
construct such a map. But first we want to comment on the spinor bundle(s)
of such a manifold. The holonomy condition means that the structure group of the
frame bundle of the oriented manifold can be reduced from SO(8) to Spin(7)
and, since Spin(7) is simply connected, the bundle group can be lifted to the
double cover Spin(8) of SO(8). Thus, Spin(7)-holonomy implies that the
manifold is Spin(8) and consequently, Spin$^c$(8).

\vskip 0.2cm
\noindent
{\bf Remark.}
The manifold $M$, being a Spin(8) manifold, possesses a real spinor bundle of
dimension 16 and a complex spinor bundle of the same dimension, which is
simply the complexification of the former. The above-mentioned complex spinor
bundle $W$ in the context of the Spin$^c$(8) structure on $M$ is nothing else
than the complex spinor bundle related to the Spin(8) structure on $M$, if the
Spin$^c$(8) structure is the one canonically associated with the Spin(8)
structure. Thus there are real spinor half-bundles $W_{r}^\pm$ with
$W^\pm = W_r \otimes \mathbf{C}$, so that we can talk about real and
imaginary parts of complex spinors.

The extension of $\Gamma$ to the even forms on $M$ leads to a
map $\rho:\Lambda^{2k}(M)\to End(W)$
given by
$$\rho(e_i^*\wedge \dots\wedge
e_{2k}^*)=\Gamma(e_1)\dots\Gamma(e_{2k}).\eqno(2.9)$$
With $\Gamma$ as given in Eq.(2.8), $\rho$ will be  block
diagonal, and one can define the maps
$$\rho^\pm(\eta)=\rho(\eta)\mid_{W^\pm},\eqno(2.10)$$
for $\eta$ in $\Lambda^{2k}(M)$.

The matrices $\gamma(e_i)=\gamma_i$ are characterized by
$$ \gamma_1=I,\quad \gamma_j^2+I=0,\quad
\gamma_j\gamma_k+\gamma_k\gamma_j=0,\quad {\rm for} \quad j\ge2,j\ne k.
\eqno(2.11)
$$
In 8-dimensions, the $\gamma_j$'s are real, skew-symetric
matrices and the set
$$
\{ \gamma_2,\gamma_3,\dots,\gamma_8,\gamma_2\gamma_3,\dots,\gamma_7\gamma_8\}
\eqno(2.12)$$
is orthonormal.
The effects of the $\rho^\pm $ maps are given by
$$
\rho^\pm(e_1^*\wedge e_j^*)= \pm \gamma_j,\quad
\rho^\pm(e_j^*\wedge e_k^*)= \gamma_j\gamma_k,$$
$$\rho^\pm(e_1^*\wedge e_j^*\wedge e_k^*\wedge e_l^*)
=\pm\gamma_j\gamma_k\gamma_l,\quad
\rho^\pm(e_j^*\wedge e_k^*\wedge e_l^*\wedge e_m^*)
=\gamma_j\gamma_k\gamma_l\gamma_m.\eqno(2.13)$$

The image of the volume form of $M$
allows to define duality identifications.
The $\gamma_j$ matrices satisfy the relations
$\gamma_2\gamma_3\gamma_4\gamma_5\gamma_6\gamma_7\gamma_8
=-I$
which allows the duality identifications
$$
\gamma_2\gamma_3\gamma_4\gamma_5\gamma_6= \gamma_7\gamma_8\quad
\gamma_2\gamma_3\gamma_4\gamma_5= -\gamma_6\gamma_7
\gamma_8.\eqno(2.14).$$

 \medskip
\noindent
{\bf The Dirac operator and the
Weitzenb\"ock formula.}
A Hermitian connection on $W$, compatible with the Levi-Civita
connection of the manifold induces an imaginary valued connection
on a certain associated
line bundle called the {\it virtual line bundle}
$L_\Gamma^{1/2}$. The
corresponding connection 1-form is denoted by $A$ and its curvature
2-form by
$F\in \Omega^2(M,i\bf R)$.
The Dirac operators $D^\pm$ corresponding to $A$ are the maps
$$D^+ : C^\infty (M,W^+)\rightarrow C^\infty (M,W^-),$$
$$D^- : C^\infty (M,W^-)\rightarrow C^\infty (M,W^+)\eqno(2.15)$$
defined by
$$D^+
(\phi^+)=\sum_{i=1}^{2n}{-\gamma^*(e_i){\nabla_{e_i}}}(\phi^+),$$
$$D^-
(\phi^-)=\sum_{i=1}^{2n}{\gamma(e_i){\nabla_{e_i}}}(\phi^-),\eqno(2.16)$$
where
$\{e_1,e_2,\cdots,{e_{2n}}\}$
is any local
orthonormal frame.

The Weitzenb\"ock formula is the key in writing the action
integral for the Seiberg-Witten equations \cite{Lawson}. One has
 $$D^-D^+\phi^+=\nabla^*\nabla\phi^+
+\ts\frac{1}{4} s\phi^+ +\rho^+(F)\phi^+,$$
 $$D^+D^-\phi^-=\nabla^*\nabla\phi^-
+\ts\frac{1}{4} s\phi^- +\rho^-(F)\phi^-,\eqno(2.17)$$
where $s$ is the scalar curvature of $M$ and
$\nabla^*$ is the $L_2$-adjoint of $\nabla$, i.e.,
$$ \int (\nabla^*\nabla\phi^\pm,\phi^\pm)dvol=
\int (\nabla\phi^\pm,\nabla\phi^\pm)dvol.\eqno(2.18)$$
Taking the inner product of Eq. (2.17)  with $\phi^\pm$
and integrating over $M$, we obtain
$$\int_M \vert D^\pm\phi^\pm\vert^2 dvol=
\int_M\left[\vert \nabla\phi^\pm\vert^2
+\ts\frac{1}{4} s \vert \phi^\pm\vert^2
+(\rho^\pm(F)\phi^\pm,\phi^\pm)\right]. \eqno(2.19)$$
Hence if $D^-\phi^-=0$,
 we have
$$
\int [\vert \nabla \phi^-\vert^2
+\ts\frac{1}{4}s\vert \phi^-\vert^2] dvol=
-\int (\rho^-(F)\phi^-,\phi^-) dvol.
\eqno(2.20)$$

\medskip
\noindent
{\bf The topological lower bound.}
The expression of the Bonan 4-form $\Phi$ in an orthonormal basis and its
eigenvectors are given in Appendix A.  The eigenvectors $\{\omega_{1j}\}$ and
$\{\omega_{jk}\}$ for $j,k=2,\dots,8$ are
a basis for local sections of $\Lambda^2(M)$, and the curvature
2-form $F$ has the splitting
$F=F^{(7)}+F^{(21)}$, where
$F^{(7)}=\sum i a_j\omega_{1j}$,
$F^{(21)}=\sum i a_{kl}\omega_{kl}$.
It can be seen that
$$F^{(7)}  \wedge F^{(21)} \wedge \Phi=0,\eqno(2.21)$$
hence
$$F\wedge F\wedge \Phi=F^{(7)}  \wedge F^{(7)} \wedge \Phi+
F^{(21)} \wedge F^{(21)}\wedge \Phi . \eqno(2.22)$$
As $F$ is pure imaginary,
$$F^{(7)} \wedge F^{(7)}  \wedge \Phi=-3*(F^{(7)} ,F^{(7)} ),$$
$$F^{(21)}\wedge F^{(21)} \wedge \Phi= *(F^{(21)},F^{(21)}).\eqno(2.23)$$
Hence
$$(F^{(21)},F^{(21)})=3(F^{(7)},F^{(7)})+*(F\wedge F \wedge \Phi).
\eqno(2.24)$$
It follows that
$$\vert F\vert^2=4\vert F^{(7)}\vert^2
+\ts\frac{2}{3}*(F\wedge F\wedge\Phi) .\eqno(2.25)$$
Note that we can also write
$$\vert F\vert^2=\ts \frac{4}{3}\vert F^{(21)}\vert^2
-\ts\frac{2}{3}*(F\wedge F\wedge\Phi) .\eqno(2.26)$$

\section{The Monopole Equations}

We have previously given \cite{BDK2} an elliptic set of monopole equations using a projection
with $\rho^+(F^{(7)})$, where we denoted $F^{(7)}$ as $F^+$,
and $\omega_{1j}$ by $f_{j-1}$.
\newpage
 The same formula (Eq.(26) in \cite{BDK2})
can be applied as well using a
projection with $\rho^-$. Hence we can write
$$\rho^\pm (F^{(7)})=\sum_{j=2}^8
\langle \rho^\pm(\omega_{1j}),\phi^\pm(\phi^\pm)^*\rangle
\vert \rho^\pm( \omega_{1j})\vert^{-2}\rho^\pm (\omega_{1j}),\eqno(3.1)$$
where $\omega_{1j}$'s are an orthonormal basis for the subspace $F^{(7)}$.
We may choose the $\omega_{1j}$'s as in (A.3).
With respect to this basis we can write
$$F^{(7)}=\sum_{j=2}^8 ia_{1j}\omega_{1j},\eqno(3.2)$$
then
$$\rho^\pm({F}^{(7)})=\sum_{j=2}^8 ia_{1j}\rho^\pm(\omega_{1j}),\eqno(3.3)$$
and the monopole equations are
$$ia_{1j}=
{\vert \rho^\pm(\omega_{1j})\vert}^{-2}
\langle \rho^\pm(\omega_{1j}),\phi^\pm (\bar{\phi}^\pm)^t\rangle,\quad\quad
j=2,\dots,8.\eqno(3.4)$$
We now discuss the effects of the maps $\rho^\pm$.
In the equations that follow, the $\omega_{1j}$'s are determined via the
Spin(7) invariant Bonan 4-form. The
first terms at the right hand sides of the equations are the
 the expressions of $\rho^\pm(\omega_{1j})$'s for an arbitrary spin$^c$
structure, while the second set of terms are the expressions corresponding to
the spin$^c$ structure given by $\gamma_j=\omega_{1j}$.  Here we remind
that there is an abuse
of notation, as $\omega_{1j}$ denotes the skew symmetric matrix corresponding
to the 2-form. Similarly in Eqs.(3.5,6), $e_{jk}$'s are skew symmetric
matrices. We use the notation $\gamma_{jk}=\gamma_j\gamma_k$. Then
 \begin{eqnarray}
\rho^+(\omega_{12})&=&\gamma_5+\gamma_{26}+\gamma_{37}+\gamma_{48}
=2[e_{13}-e_{24}-e_{57}-e_{68}]\nonumber\\
\rho^+(\omega_{13})&=&\gamma_2+\gamma_{34}-\gamma_{56}-\gamma_{78}
=2[e_{15}-e_{26}+e_{37 } +e_{48}]\nonumber\\
\rho^+(\omega_{14})&=&\gamma_6-\gamma_{25}-\gamma_{38}+\gamma_{47}
=2[e_{17}+e_{28}-e_{35 } +e_{46}]\nonumber\\
\rho^+(\omega_{15})&=&\gamma_3-\gamma_{24}-\gamma_{57}+\gamma_{68}
=2[e_{12}+e_{34}+e_{56 } -e_{78}]\nonumber\\
\rho^+(\omega_{16})&=&\gamma_7+\gamma_{28}-\gamma_{35}-\gamma_{46}
=2[e_{14}+e_{23}-e_{58 } +e_{67}]\nonumber\\
\rho^+(\omega_{17})&=&\gamma_4+\gamma_{23}-\gamma_{58}-\gamma_{67}
=2[-e_{16}-e_{25}-e_{3 8 }+e_{47}]\nonumber\\
\rho^+(\omega_{18})&=&\gamma_8-\gamma_{27}+\gamma_{36}-\gamma_{45}
=2[e_{18}-e_{27}-e_{36 } -e_{45}]. \nonumber
\end{eqnarray}
$$\eqno(3.5)$$

\begin{eqnarray}
\rho^-(\omega_{12})&=&
-\gamma_5+\gamma_{26}+\gamma_{37}+\gamma_{48}=-4 e_{68},\nonumber\\
\rho^-(\omega_{13})&=&
-\gamma_2+\gamma_{34}-\gamma_{56}-\gamma_{78}=-4e_{26},\nonumber\\
\rho^-(\omega_{14})&=&
-\gamma_6-\gamma_{25}-\gamma_{38}+\gamma_{47}= 4e_{46},\nonumber\\
\rho^-(\omega_{15})&=&
-\gamma_3-\gamma_{24}-\gamma_{57}+\gamma_{68}= 4e_{56},\nonumber\\
\rho^-(\omega_{16})&=&
-\gamma_7+\gamma_{28}-\gamma_{35}-\gamma_{46}= 4e_{67},\nonumber\\
\rho^-(\omega_{17})&=&
-\gamma_4+\gamma_{23}-\gamma_{58}-\gamma_{67}=-4e_{16},\nonumber\\
\rho^-(\omega_{18})&=&
-\gamma_8-\gamma_{27}+\gamma_{36}-\gamma_{45}=-4e_{36}.\nonumber
\end {eqnarray}
$$\eqno(3.6)$$

We shall not give the explicit expression of the
$\rho^-(\omega_{jk})$'s, but we indicate the method of computation.
 As an example,
we compute $\rho^-(\omega)$
for $\omega=e_{12}+a e_{34}+b e_{56}+c e_{78}$.  For $1+a-b-c=0$, $\omega$
belongs to the $21$ dimensional  subspace.
We have chosen the labeling of the $\omega_{jk}$'s so that
if $\gamma_j=\omega_{1j}$, then the matrix
$\gamma_{jk}=\gamma_j\gamma_k$
is the skew symmetric matrix corresponding to the 2-form
$\omega_{jk}$.
Then using Eqs.(A.3-4) we have
$$\rho^-(\omega)=-\gamma_2+a\gamma_{34}+b\gamma_{56}+c\gamma_{78}$$
$$=-  (e_{15}+e_{26}+e_{37}+e_{48})
   -a (e_{15}+e_{26}-e_{37}-e_{48})
   -b (e_{15}-e_{26}+e_{37}-e_{48})
   -c (e_{15}-e_{26}-e_{37}+e_{48})
$$
$$
=-( 1+a+b+c)e_{15}
 +(-1-a+b+c)e_{26}
 +(-1+a-b+c)e_{37}
 +(-1+a+b-c)e_{48}.\eqno(3.7)$$
Note that for $\omega$ in the $21$ dimensional subspace,
the coefficient of $e_{26}$ is zero.
With similar computations it can be seen that the coefficients of
$e_{j6}$ vanish for each $j$, whenever $\omega $ belongs to
the 21 dimensional space.
It follows that the 6th row and the 6th column of $\rho^-(F^{(21)})$ are
zero. This will be the key fact in constructing the action integral.

For the specific spin$^c$ structure above,
our monopole equations  corresponding to the coupling to a positive
spinor \cite{BDK2} are
$$
4a_{12}=
F_{15}+F_{26}+F_{37}+F_{48}=
\ts \frac{1}{4}(\phi_1 {\bar \phi }_3- \phi_3{\bar \phi }_1
-\phi_2 {\bar \phi }_4+\phi_4 {\bar \phi }_2
-\phi_5 {\bar \phi }_7+\phi_7 {\bar \phi }_5
-\phi_6 {\bar \phi }_8+\phi_8 {\bar \phi }_6),$$
$$
4a_{13}=
F_{12}+F_{34}-F_{56}-F_{78}=
\ts \frac{1}{4}(\phi_1 {\bar \phi }_5- \phi_5
{\bar \phi
}_1 -\phi_2 {\bar \phi }_6+\phi_6 {\bar \phi }_2
+\phi_3 {\bar \phi }_7-\phi_7 {\bar \phi }_3
+\phi_4 {\bar \phi }_8-\phi_8 {\bar \phi }_4),$$
$$
4a_{14}=
F_{16}-F_{25}-F_{38}+F_{47}=
\ts \frac{1}{4}(\phi_1 {\bar \phi }_7- \phi_7
{\bar \phi
}_1 +\phi_2 {\bar \phi }_8-\phi_8 {\bar \phi }_2
-\phi_3 {\bar \phi }_5+\phi_5 {\bar \phi }_3
+\phi_4 {\bar \phi }_6-\phi_6 {\bar \phi }_4),$$
$$
4a_{15}=
F_{13}-F_{24}-F_{57}+F_{68}=
\ts \frac{1}{4}(\phi_1 {\bar \phi }_2- \phi_2
{\bar \phi
}_1 +\phi_3 {\bar \phi }_4-\phi_4 {\bar \phi }_3
+\phi_5 {\bar \phi }_6-\phi_6 {\bar \phi }_5
-\phi_7 {\bar \phi }_8+\phi_8 {\bar \phi }_7),$$
$$
4a_{16}=
F_{17}+F_{28}-F_{35}-F_{46}=
\ts \frac{1}{4}(\phi_1 {\bar \phi }_4- \phi_4
{\bar \phi
}_1 +\phi_2 {\bar \phi }_3-\phi_3 {\bar \phi }_2
-\phi_5 {\bar \phi }_8+\phi_8 {\bar \phi }_5
+\phi_6 {\bar \phi }_7-\phi_7 {\bar \phi }_6),$$
$$
4a_{17}=
F_{14}+F_{23}-F_{58}-F_{67}=
\ts \frac{1}{4}(-\phi_1 {\bar \phi }_6+ \phi_6
{\bar \phi
}_1 -\phi_2 {\bar \phi }_5+\phi_5 {\bar \phi }_2
-\phi_3 {\bar \phi }_8+\phi_8 {\bar \phi }_3
+\phi_4 {\bar \phi }_7-\phi_7 {\bar \phi }_4),$$
$$
4a_{18}=
F_{18}-F_{27}+F_{36}-F_{45}=
\ts \frac{1}{4}(\phi_1 {\bar \phi }_8- \phi_8
{\bar \phi
}_1 -\phi_2 {\bar \phi }_7+\phi_7 {\bar \phi }_2
-\phi_3 {\bar \phi }_6+\phi_6 {\bar \phi }_3
-\phi_4 {\bar \phi }_5+\phi_5 {\bar \phi }_4),$$
$$\eqno(3.8)$$
while for the coupling to a negative spinor here we give
the monopole equations
$$
4a_{12}=
F_{15}+F_{26}+F_{37}+F_{48}=
\ts \frac{1}{2}
(-\phi_6 {\bar \phi }_8+\phi_8 {\bar \phi }_6),$$
$$
4a_{13}=
F_{12}+F_{34}-F_{56}-F_{78}=
\ts \frac{1}{2}
(-\phi_2 {\bar \phi }_6+\phi_6 {\bar \phi }_2 ),$$
$$
4a_{14}=
F_{16}-F_{25}-F_{38}+F_{47}=
\ts \frac{1}{2}
(\phi_4 {\bar \phi }_6-\phi_6 {\bar \phi }_4),$$
$$
4a_{15}=
F_{13}-F_{24}-F_{57}+F_{68}=
\ts \frac{1}{2}
(\phi_5 {\bar \phi }_6-\phi_6 {\bar \phi }_5),$$
$$
4a_{16}=
F_{17}+F_{28}-F_{35}-F_{46}=
\ts \frac{1}{2}
(\phi_6 {\bar \phi }_7-\phi_7 {\bar \phi }_6),$$
$$
4a_{17}=
F_{14}+F_{23}-F_{58}-F_{67}=
\ts \frac{1}{2}
(-\phi_1 {\bar \phi }_6+ \phi_6{\bar \phi}_1 ),$$
$$
4a_{18}=
F_{18}-F_{27}+F_{36}-F_{45}=
\ts \frac{1}{2}
(-\phi_3 {\bar \phi }_6+\phi_6 {\bar \phi }_3),$$
$$\eqno(3.9)$$

\section{The Action Integral}

We prove now that the monopole equations (3.9) for $F^{(7)}$ and  $\phi^-$
can be derived from an action principle.  However, this is not possible
for any $\phi^-$,   but for a  special one given by Eq.(4.1) below. The
subtle point here is the choice of a real section of $W^-$,
which is possible as discussed
in the "Remark" in Section 2.

We recall that $M$ is an 8-manifold with Spin(7) holonomy, with scalar
curvature $s$ and Bonan 4-form $\Phi$. Given any spin$^c$ structure $(\Gamma,
W)$, the subspace $W^-$ and the map $\rho^-$ are defined by (2.7) and (2.10),
$A$ is a $U(1)$ connection on the ``virtual line bundle" $L_\Gamma^{1/2}$
(as in Section 2) and $F$ is the curvature of $A$. The Bonan 4-form determines
an orthogonal direct sum decomposition of the 2-form $F$ by the map (A.2), and
$F^{(7)}$ and $F^{(21)}$ are the projections of $F$ onto these subspaces.

\medskip
\noindent
{\bf Proposition 4.1.} Let $M$, $F$ and $\Phi$ be as above and define a section
$\phi$ of $W^-$ by
$$\phi=(1-P)U+iPU\eqno(4.1) $$
where $U$ is a real section of $W^-$ and
$$P=\ts \frac{1}{8}-\ts \frac {1}{16}\rho^-(\Phi).\eqno(4.2)$$
Then the monopole equations
$$D^-\phi=0,\quad
\rho^-(F^{(7)})=\ts \frac{1}{2}(\phi\bar{\phi}^t-\bar{\phi}\phi^t),\quad
{\rm div}A=0\eqno(4.3)$$
are absolute minimizers of the action
$$
I(A,U)=\int_M\left[ \vert F\vert^2
                  + \vert \nabla \phi\vert^2
+\ts\frac{1}{4}s\vert \phi\vert^2
+\vert \phi\bar{\phi}^t-\bar{\phi}\phi^t\vert^2\right] dvol
\eqno(4.4)
$$
and $ I(A,U)\ge \frac{2}{3}\int_M F\wedge F\wedge \Phi$.

\medskip
\noindent
{\it Proof.}  We first use Eqs. (2.25) and (2.20) to express the action as
$$I(A,U)=
\ts \frac{2}{3}\int_M F\wedge F\wedge \Phi
+\int_M  [   4 \vert F^{(7)}\vert^2
-(\rho^-(F)\phi,\phi)
+\ts\frac{1}{4}s\vert \phi\vert^2
+\vert \phi\bar{\phi}^t-\bar{\phi}\phi^t\vert^2      ] dvol
\eqno(4.5)
$$
Next we show that $\vert F^{(7)}\vert^2=\vert \rho^-(F^{(7)})\vert^2$.
For this note that $\rho^-(\omega_{1j})$ is of the form
$-\gamma_i+\gamma_j\gamma_k+\gamma_m\gamma_n+\gamma_p\gamma_q$ (see Eqs.(3.6))
with distinct set of indices. The orthonormality of the set (2.12) implies that
$\vert \rho^-(\omega_{1j})\vert^2=4$.  On the other hand
from (A.3), $\vert \omega_{1j} \vert^2=4$, hence the $\rho^-$ map is an
isometry. It follows that
$$I(A,U)=
\ts \frac{2}{3}\int_M F\wedge F\wedge \Phi
+\int_M\left[ 4 \vert \rho^-(F^{(7)})\vert^2
-(\rho^-(F)\phi,\phi)
+\ts\frac{1}{4}s\vert \phi\vert^2
+\vert \phi\bar{\phi}^t-\bar{\phi}\phi^t\vert^2\right] dvol
\eqno(4.6)
$$
The main difficulty is to show that $(\rho^-(F^{(21)})\phi,\phi)=0$ for $\phi$
given by Eq. (4.1). For this we first prove the following Lemma.

\medskip
\noindent
{\bf Lemma 4.2.}  Let $P$ be as in Eq. (4.2). Then
$$\rho^-(F^{(21)})P=0,\quad
\rho^-(F^{(7)})P+P\rho^-(F^{(7)})=\rho^-(F^{(7)}).\eqno(4.7)$$
\medskip
\noindent
{\it Proof of Lemma 4.2.} Let $Q= -\frac{1}{2} \rho^-(\Phi)$.  For
$\Phi$ given by Eq.(A.1), $Q$ is the symmetric matrix given by
$$Q =
\gamma_2\gamma_3\gamma_4
  -\gamma_2\gamma_5\gamma_6
  -\gamma_2\gamma_7\gamma_8
  -\gamma_3\gamma_5\gamma_7
  +\gamma_3\gamma_6\gamma_8
  -\gamma_4\gamma_5\gamma_8
  -\gamma_4\gamma_6\gamma_7 .
  \eqno(4.8)
$$
Using the duality relations (2.14), it can be seen that $Q^2=7+6Q$, hence the
minimal polynomial of $Q$ is
$$(Q-7I)(Q+I)=0.\eqno(4.9)$$
Let the dimension of the eigenspaces corresponding to the eigenvalues $7$ and
$-1$ be $m$ and $8-m$.  The orthogonality of the set (2.12)
implies that $Q$ is
traceless. On the other hand, $trQ=7m+(-1)(8-m)=0$, hence $m=1 $. If we set
$P=\frac{1}{8}(Q+I)$, then   $P$ has rank 1 and
$$P^2=P,\eqno(4.10)$$
hence $P$ is a projection onto a 1-dimensional subspace of $W^-$. To prove
Eq.(4.7) we need to compute the relevant expressions for each basis element.
We give the details of the computation for
$$\omega=e_{12}+ae_{34}+b e_{56}+c e_{78}\eqno(4.11)$$
which will prove the result for $\omega_{13}$,  $\omega_{24}$,
$\omega_{68}$ and  $\omega_{75}$.  We rewrite $(Q+I)$ four times with
appropriate duality identifications and multiply with $\omega$.
We use below the notation $\gamma_{ijk}=\gamma_i\gamma_j \gamma_k$.
\begin{eqnarray}
(Q+I)\omega&=&
(I+\gamma_{234}- \gamma_{256} - \gamma_{278}
+  \gamma_{2468} - \gamma_{2457} + \gamma_{2367}+  \gamma_{2358} )
(- \gamma_{2})
\nonumber \\
&&(I+\gamma_{234}+ \gamma_{3478}+  \gamma_{3456}
- \gamma_{357} + \gamma_{368} - \gamma_{458} - \gamma_{467})(a\gamma_{34})
\nonumber \\
&&(I-\gamma_{5678}- \gamma_{256}+  \gamma_{3456}
- \gamma_{357} + \gamma_{368} - \gamma_{458} - \gamma_{467})(a\gamma_{56})
\nonumber \\
&&(I-\gamma_{5678}+ \gamma_{3478} - \gamma_{278}
- \gamma_{357} + \gamma_{368} - \gamma_{458} - \gamma_{467})(a\gamma_{78})
\nonumber\\
&=&(1+a-b-c)(-\gamma_2+\gamma_{34}-\gamma_{56}-\gamma_{78}
- \gamma_{468} + \gamma_{457} - \gamma_{367} + \gamma_{358}).
\nonumber 
\end{eqnarray}
$$\eqno(4.12)$$
If $\omega$ belongs to the $21$ dimensional subspace then $1+a-b-c=0$, and hence
$(Q+I)\rho^-(\omega)=0$. On the other hand if $\omega$ belongs to the
7-dimensional subspace, $(Q+I)\rho^-(\omega)\ne 0$, but it can be seen that
$$(Q+I)\rho^-(\omega)+\rho^-(\omega)(Q+I)=2(1+a-b-c)
(-\gamma_2+\gamma_{34} +\gamma_{56} +\gamma_{78}) \eqno(4.13)$$
and as $a=1$, $b=c=-1$, it follows that
$(Q+I)\rho^-(\omega)+\rho^-(\omega)(Q+I)=8 \rho^-(\omega)$ and we obtain Eq.
(4.7).

\medskip
\noindent
{\bf Remark 4.3.} For the specific choice of the spin$^c$ structure
corresponding to $\gamma_j=\omega_{1j}$, $P$ is the diagonal matrix with the
only nonzero entry, $P_{66}=1$.  From the remark after Eq.(3.7) we see that the
6th row and the 6th column of $\rho^-(\omega_{jk})$'s are zero while in
$\rho^-(\omega_{1j})$'s non-zero elements are only in the 6th row and the 6th
column.

\medskip
We now continue with the proof of Proposition 4.1.  First note that as $F$ is
pure imaginary and the representation is real, $\rho^-(F)$ is hermitian but
skew-symmetric.  Now using $\rho^-(F^{(21)})P=0$, we have
$$(\rho^-(F^{(21)})\phi,\phi)=(\rho^-(F^{(21)})U,(1-P)U+iPU)
=(U,\rho^-(F^{(21)})U)=0.\eqno(4.14)$$
where the second equality follows from hermiticity  and the third from
skew-symmetry.

Then from Eq.(2.3),
$$(\rho^-(F^{(7)})\phi,\phi)=4(\rho^-(F^{(7)}),
\phi\bar{\phi}^t-\bar{\phi}\phi^t)\eqno(4.15)$$
and Eq.(4.6) reduces to
$$I(A,U)=
\ts \frac{2}{3}\int_M F\wedge F\wedge \Phi
+\int_M\left[ 4 \vert \rho^-(F^{(7)})\vert^2
-4 (\rho-(F^{(7)}),\phi\bar{\phi}^t-\bar{\phi}\phi^t)
+\vert \phi\bar{\phi}^t-\bar{\phi}\phi^t\vert^2\right] dvol
\eqno(4.16)
$$
which can be written as
$$I(A,U)=
\ts \frac{2}{3}\int_M F\wedge F\wedge \Phi
+\int_M  \vert 2\rho^-(F^{(7)})
               -(\phi\bar{\phi}^t-\bar{\phi}\phi^t)\vert^2 dvol
\eqno(4.17)
$$
which gives the monopole equations.

As the left hand side of the matrix equation $\rho^-(F^{(7)})=\frac{1}{2}
(\phi\bar{\phi}^t-\bar{\phi}\phi^t)$ belongs to a 7-dimensional subspace, we
need to check its compatibility. Given any skew-symmetric matrix $A$, there is
an orthogonal  decomposition
$$A=(PA+AP)+(I-P)A(I-P).\eqno(4.18)$$
The orthogonality of the summands can be seen from
$$tr((PA+AP)(I-P)A(I-P))=tr(PA(I-P)A(I-P))=tr(A(I-P)A(I-P)P)=0.$$
From Eq.(4.7), $\rho^-(F^{(7)})=P\rho^-(F^{(7)})+\rho^-(F^{(7)})P$. Hence we need
to show that the right hand side satisfies the same relation. It can be seen
that
$$\ts \frac{1}{2}(\phi\bar{\phi}^t-\bar{\phi}\phi^t)
=\ts \frac{1}{2}[-i(I-P)UU^tP+iPUU^t(I-P)]=i[-UU^tP+PUU^t].\eqno(4.19)$$
But
$$P[PUU^t-UU^tP]+[PUU^t-UU^tP]P=PUU^t-UU
^tP , \eqno(4.20)$$
so both sides of the equation belong
to the same linear subspace. This
completes the proof of Proposition 4.1.

\medskip

For $\gamma_j=\omega_{1j}$, the explicit form of the monopole equations can be
obtained by putting
$$\bar{\phi}_6=-\phi_6,\quad
  \bar{\phi}_j=\phi_j,\quad {\rm for}\quad j\ne 6\eqno(4.21)$$
in Eqs.(3.9).

Finally, we shall show that the new set of equations are elliptic.
In a system of $n$ first order partial differential equations in the
independent variables $x_1,\dots , x_k$ and the dependent variables $u_1,\dots
,u_n$, the replacement of $\frac {\partial u_i}{\partial x_j}$ by $\xi_ju_i$,
where $\xi_j$'s are indeterminates, one obtains (see e.g. \cite{John}) a linear homogeneous
system of equations for the $u_i$'s. The characteristic determinant of the system is the
determinant of  the coeficient matrix of the corresponding linear system.
The system of partial differential equations is called elliptic, if the
characteristic determinant has no real roots. The ellipticity property of a
system is independent of its inhomogeneous part.

\medskip
\noindent
{\bf Proposition 4.4.}  The system of
equations (4.3) is elliptic.

\medskip
\noindent
{\it Proof.}  It is known that the
system  of equations $F^{(7)}=0$ together with the Coulomb gauge condition
$div(A)=0$ are elliptic [9].  If $F^{(7)}=\sum_{j=2}^8 ia_j\omega_{1j}$, as
$\rho^-(\omega_{1j})$'s are linearly independent, $\rho^-(F^{(7)})=0$ if and
only if $F^{(7)}=0$.  Thus the inhomogeneous system
 $\rho^-(F^{(7)})=\frac{1}{2}
(\phi\bar{\phi}^t-\bar{\phi}\phi^t)$ together with the Coulomb gauge condition
is elliptic.
To prove the ellipticity of the system $D^-\phi=0$, let   $\phi=SU$ and
$K$ be the characterisitic determinant of the system $D^-U=0$. It can be
seen that the characteristic determinant of the system $D^-(SU)=0$ is
$\tilde{K}=Kdet(S)$.
Hence we need to show that $S$ is nonsingular. Using the minimal polynomial of
$Q$ it can be checked that
the minimal polynomial of $S$ is
$$S^2-(1+i)S+i $$
which shows that it is nonsingular and completes the proof.

\newpage

\section{Appendix A: }

We recall that, in an orthonormal basis $\{e_i\}$, $i=1,\dots ,8$,  the Bonan
4-form $\Phi$ can be written as

\begin{eqnarray}
\Phi&=&
[e_{1234}-e_{1256}-e_{1278}-e_{1357}+e_{1368}
-e_{1458}-e_{1467}\nonumber\\
&&+e_{5678}-e_{3478}-e_{3456}-e_{2468}+e_{2457}-e_{2367}-e_{2358}].\nonumber
\end{eqnarray}
$$ \eqno(A.1)$$
where $e_{ijkl}=e^*_i\wedge e^*_j \wedge e^*_k \wedge e^*_l $.
The form $\Phi$ is Spin(7) invariant, hence it can be extended to the manifold.
$\Phi$ is self-dual and it defines a linear map on 2-forms as
$$\omega\to *(\Phi\wedge \omega),\eqno(A.2)$$
with eigenvalues $3$ and $-1$. This map has 7 and 21 dimensional eigenspaces
and a basis consisting of the corresponding  eigenvectors are given below.

\medskip
\noindent
{\it Eigenvectors corresponding to the eigenvalue $3$:}
\begin{eqnarray}
\omega_{12}=&e_{15}+e_{26}+e_{37}+e_{48}\nonumber\\
\omega_{13}=&e_{12}+e_{34}-e_{56}-e_{78}\nonumber\\
\omega_{14}=&e_{16}-e_{25}-e_{38}+e_{47}\nonumber\\
\omega_{15}=&e_{13}-e_{24}-e_{57}+e_{68}\nonumber\\
\omega_{16}=&e_{17}+e_{28}-e_{35}-e_{46}\nonumber\\
\omega_{17}=&e_{14}+e_{23}-e_{58}-e_{67}\nonumber\\
\omega_{18}=&e_{18}-e_{27}+e_{36}-e_{45}\nonumber
\end{eqnarray}
$$\eqno(A.3)$$
\medskip

The labeling of the eigenvectors corresponding to the eigenvalue $-1$ is
done as follows.  By abuse of notation, let $\omega_{1j} $
be the skew-symmetric
matrix corresponding to the 2-form $\omega_{1j}$. Then $\omega_{jk}$ is the
2-form corresponding to the matrix $\omega_{1j}\omega_{1k}$.

\newpage
\noindent
{\it Eigenvectors corresponding to the eigenvalue $-1$:}
\begin{eqnarray}
\omega_{87}=&e_{15}-e_{26}-e_{37}+e_{48}\nonumber\\
\omega_{68}=&e_{12}-e_{34}+e_{56}-e_{78}\nonumber\\
\omega_{67}=&e_{16}+e_{25}+e_{38}+e_{47}\nonumber\\
\omega_{84}=&e_{13}+e_{24}+e_{57}+e_{68}\nonumber\\
\omega_{74}=&e_{17}-e_{28}+e_{35}-e_{46}\nonumber\\
\omega_{46}=&e_{14}-e_{23}+e_{58}-e_{67}\nonumber\\
\omega_{72}=&e_{18}+e_{27}-e_{36}-e_{45}\nonumber\\
\nonumber&\\
\omega_{43}=&e_{15}+e_{26}-e_{37}-e_{48}\nonumber\\
\omega_{75}=&e_{12}+e_{34}+e_{56}+e_{78}\nonumber\\
\omega_{32}=&e_{16}-e_{25}+e_{38}-e_{47}\nonumber\\
\omega_{37}=&e_{13}-e_{24}+e_{57}-e_{68}\nonumber\\
\omega_{83}=&e_{17}+e_{28}+e_{35}+e_{46}\nonumber\\
\omega_{53}=&e_{14}+e_{23}+e_{58}+e_{67}\nonumber\\
\omega_{36}=&e_{18}-e_{27}-e_{36}+e_{45}\nonumber\\
&\nonumber\\
\omega_{65}=&e_{15}-e_{26}+e_{37}-e_{48}\nonumber\\
\omega_{24}=&e_{12}-e_{34}-e_{56}+e_{78}\nonumber\\
\omega_{58}=&e_{16}+e_{25}-e_{38}-e_{47}\nonumber\\
\omega_{26}=&e_{13}+e_{24}-e_{57}-e_{68}\nonumber\\
\omega_{52}=&e_{17}-e_{28}-e_{35}+e_{46}\nonumber\\
\omega_{28}=&e_{14}-e_{23}-e_{58}+e_{67}\nonumber\\
\omega_{45}=&e_{18}+e_{27}+e_{36}+e_{45}\nonumber
\end{eqnarray}
$$\eqno(A.4)$$

\end{document}